\documentclass[aps,prb,twocolumn,floatfix,superscriptaddress]{revtex4}
\usepackage{times}
\usepackage{graphicx}
\usepackage{epsfig}
\usepackage{amsmath}
\usepackage{amssymb}
\usepackage[version=3]{mhchem}
\usepackage{nicefrac,xfrac}

\begin{document}

\title{Structural complexity in Prussian blue analogues}

\author{John Cattermull}
\affiliation{Department of Chemistry, University of Oxford, Inorganic Chemistry Laboratory, South Parks Road, Oxford OX1 3QR, U.K.}
\affiliation{Department of Materials, University of Oxford, Parks Road, Oxford OX1 3PH, U.K.}

\author{Mauro Pasta$^\ast$}
\affiliation{Department of Materials, University of Oxford, Parks Road, Oxford OX1 3PH, U.K.}

\author{Andrew L. Goodwin$^\ast$}
\affiliation{Department of Chemistry, University of Oxford, Inorganic Chemistry Laboratory, South Parks Road, Oxford OX1 3QR, U.K.}

\date{\today}
\begin{abstract}
We survey the most important kinds of structural complexity in Prussian blue analogues, their implications for materials function, and how they might be controlled through judicious choice of composition. We focus on six particular aspects: octahedral tilts, A-site `slides', Jahn--Teller distortions, A-site species and occupancy, hexacyanometallate vacancies, and framework hydration. The promising K-ion cathode material K$_x$Mn[Fe(CN)$_6$]$_y$ serves as a recurrent example that illustrates many of these different types of complexity. Our article concludes with a discussion of how the interplay of various distortion mechanisms might be exploited to optimise the performance of this and other related systems, so as to aid in the design of next-generation PBA materials.
\end{abstract}


\maketitle

\section{Introduction}

Since their crystal structure was solved half a century ago,\cite{Ludi1970} Prussian blue analogues (PBAs) have been the focus of many different areas of research; \emph{e.g.}\ as battery materials,\cite{Hurlbutt2018} molecular magnets,\cite{Verdaguer_2005} gas storage media,\cite{Kaye2005} and more.\cite{Matos-Peralta2020} In many ways this functional diversity has its roots in the compositional and structural versatility of the PBA family. PBAs are `hybrid' double perovskites (general formula {\sf A}$_2${\sf BB}$^\prime${\sf X}$_6$), in which transition metals {\sf P} and {\sf R} occupy the alternating {\sf B}- and {\sf B}$^\prime$-sites of a simple cubic lattice, are connected by cyanide linkers on the {\sf X}-site to form a framework structure, with {\sf A}-site cations and water occupying the pore-space within.\cite{Bostrom2021} A significant feature of PBA chemistry is the incorporation of hexacyanometallate, [{\sf R}$^n$(CN)$_6$]$^{(6-n)-}$, vacancies in the structure. These vacancies, along with variable oxidation states of the {\sf P}- and {\sf R}-site metals, are electronically balanced by intercalation of {\sf A}-site cations. So the universal composition is given as {\sf A}$_x${\sf P}$^{m}$[{\sf R}$^{n}$(CN)$_6$]$_{y}$\,$\cdot$\,$z$H$_2$O, subject to the charge-balance constraint $x+m=y(6-n)$. Hereafter we will use the shorthand notation {\sf A}$_x${\sf P}[{\sf R}]$_{y}$.

The crystallographic details of PBAs are known to have a significant effect on functionality; this is particularly true in the case of battery materials.\cite{Brant2019} By way of example, potassium manganese hexacyanoferrate -- a promising K-ion cathode material favoured for its high redox potential and low cost\cite{Dhir2020} --- undergoes two structural phase transitions during every charge/discharge cycle.\cite{Bie2017} These transitions are thought to impact mechanical stability, resulting in capacity loss.\cite{Bie2017} Hence, a good understanding of the structural complexities at a fundamental level will be vital for the design of next generation PBA materials, capable of competing with current state of the art lithium-ion technologies, for example.

\begin{figure}
	\centering
	\includegraphics{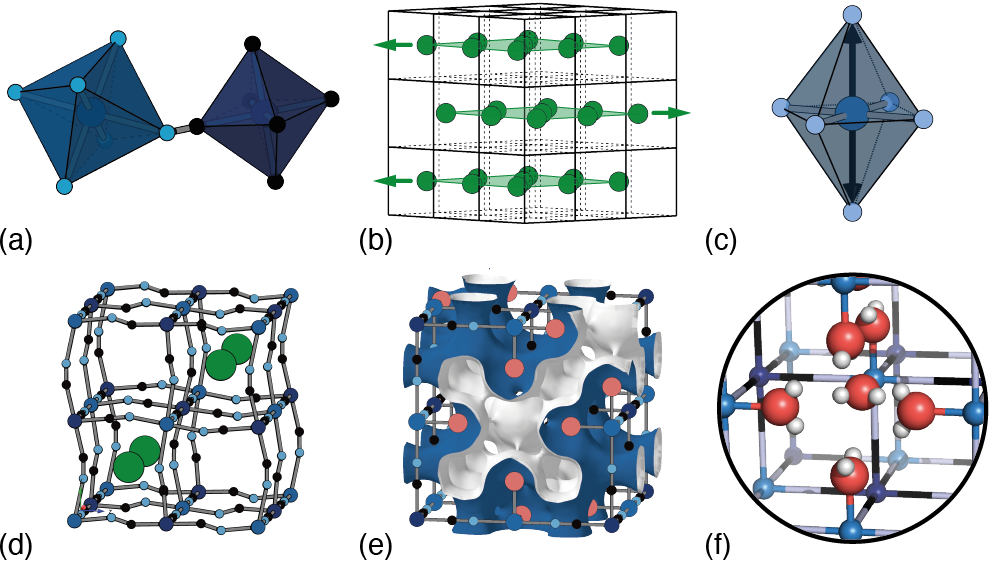}
	\caption{Common types of structural complexity in PBAs: (a) octahedral tilts, (b) correlated `slides' of A-site cations, (c) Jahn--Teller distortions, (d) A-site occupancy (dis)order, (e) hexacyanometallate vacancies and the resulting connected pore-network structure, and (f) framework hydration.}
	\label{fig1}
\end{figure}

In this Focus article we survey the key types of structural complexity in PBAs, and explore how they might be controlled through judicious choice of composition. The characteristic parent \emph{fcc} structure --- so famously associated with PBAs --- is common,\cite{vanBever1938} but there are many distortions that can break the cubic symmetry, locally or globally,\cite{Her2010} and here we will discuss the most important of these: namely, octahedral tilts, {\sf A}-site `slides', Jahn-Teller activity, cation and vacancy order, and hydration [Fig.~\ref{fig1}]. In our discussion we will address the experimental sensitivity to these various degrees of freedom, and where possible relate our observations to the specific example of potassium manganese hexacyanoferrate, K$_x$Mn[Fe]$_y$.

\section{Octahedral tilts}

\begin{figure}
	\centering
	\includegraphics[width=\columnwidth]{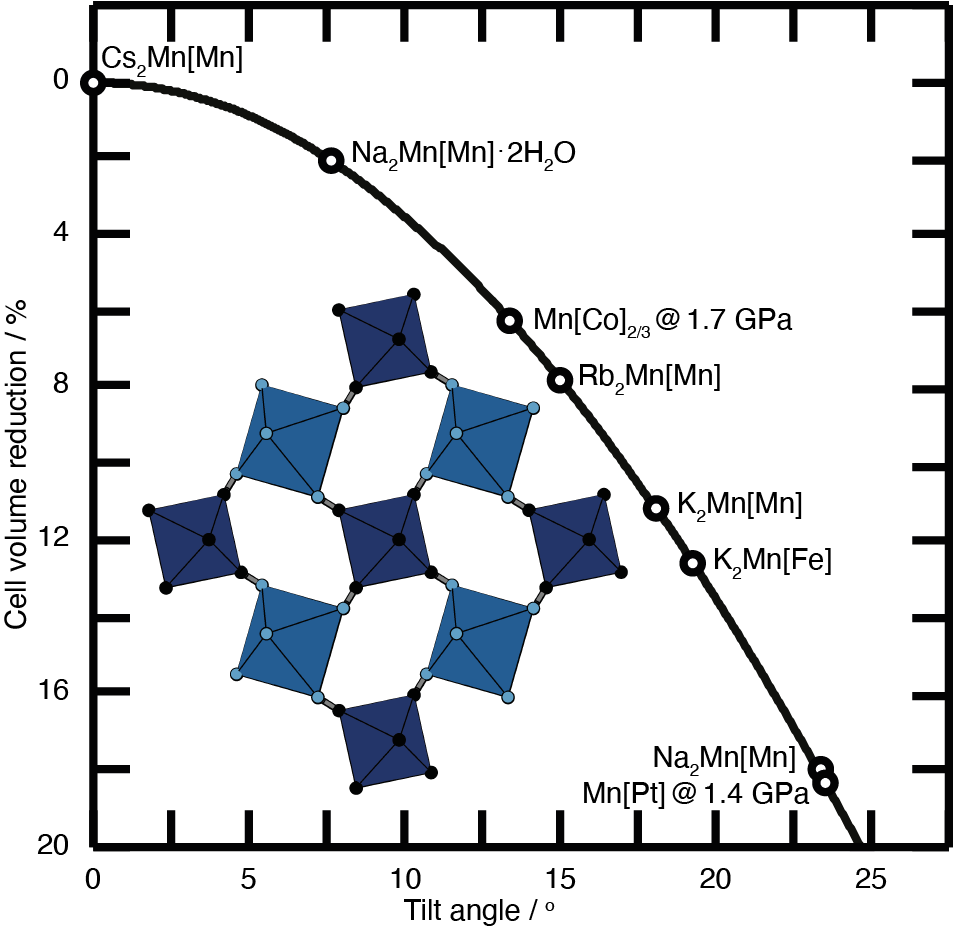}
	\caption{Approximate link between magnitude of correlated tilts and volume reduction in the PBA structure type. The solid line illustrates the geometric result for volume reduction $-\Delta V/V\simeq6r[1-\cos(\phi)]$, where the `tilt angle' $\phi$ is the average distortion of {\sf P}$\ldots${\sf R}--C and {\sf R}$\ldots${\sf P}--N angles and $r$ is the average ratio of the {\sf P}--N/{\sf R}--C bond lengths to the {\sf P}$\ldots${\sf R} distance.\cite{Her2010,Kareis2012,Bostrom2019,Fiore2020} Included in the bottom left is an illustration of one representative tilt distortion. We include the positions of a variety of PBAs discussed in the text. Note, for example, that decreasing A-site cation radius favours larger tilts, as seen for the family A$_2$Mn[Mn] (A = Cs, Rb, K, Na).}
	\label{tilts}
\end{figure}

Perhaps the most fundamental of the structural degrees of freedom in PBAs are the octahedral tilts, which are widely studied in perovskites more broadly.\cite{Glazer1972,Howard1998,Howard2004,Howard2005,Duyker2016,Bostrom2020c} Tilts are volume reducing distortions, since by bending the {\sf R}--C$\equiv$N and {\sf P}--N$\equiv$C bonds, the overall distance between neighbouring transition metals is reduced, as shown in Fig. \ref{tilts}. They can be activated by external pressure\cite{Bostrom2019} or by internal (chemical) pressure.\cite{Sugimoto2017}

The tilting of one octahedron necessarily affects that of its neighbours, and at a critical external pressure, these tilts become cooperative in a long-range sense, reducing the crystal symmetry --- usually to rhombohedral. This is seen, for example, in both stoichiometric Mn[Pt] and defective Mn[Co]$_{\nicefrac{2}{3}}$.\cite{Bostrom2019} The crucial difference observed between the two is that Mn[Co]$_{\nicefrac{2}{3}}$ shows a greater volume contraction prior to long-range symmetry breaking. Vacancies in a PBA system reduce connectivity, allowing volume contraction to originate from local distortions --- rather than involving large domains --- with columns of octahedra able to compress into the vacancies.\cite{Bostrom2016} Vacancies also increase the free pore space, hence enhancing compressibility.\cite{Bostrom2016}

Substantial tilts $>15^{\circ}$ are seen in PBA systems with high alkali-metal ion content, such as K$_2$Mn[Fe].\cite{Fiore2020} Tilts reduce the size of the {\sf A}-site cavity, and in doing so reduce the distance between the intercalated cation and the anionic cyanide linkers. In this way one might intuitively expect smaller, more polarising cations to more readily activate tilts. This is precisely what one finds in the series of manganese hexacyanomanganate systems {\sf A}$_2$Mn[Mn] ({\sf A} = Na, K, Rb, Cs) [Fig. \ref{tilts}].\cite{Her2010,Kareis2012}

The fact that tilts are activated so easily suggests they are the lowest-energy volume-reducing mode accessible to the PBA structure type. This is consistent with our understanding of the phonon spectrum of PBAs\cite{Bostrom2016} and of their role in negative thermal expansion.\cite{Goodwin2005,Chapman2006} In principle, PBAs --- even more so than double-perovskites\cite{Howard2003} --- support a large variety of different tilt distortions. But it is the very simplest tilts (\emph{e.g.}\ the rhombohedral distortion discussed above) that prevail in practice.\cite{Bostrom2019, Her2010, Fiore2020}

\section{A-site slides}
\begin{figure*}
	\centering
	\includegraphics[width=\textwidth]{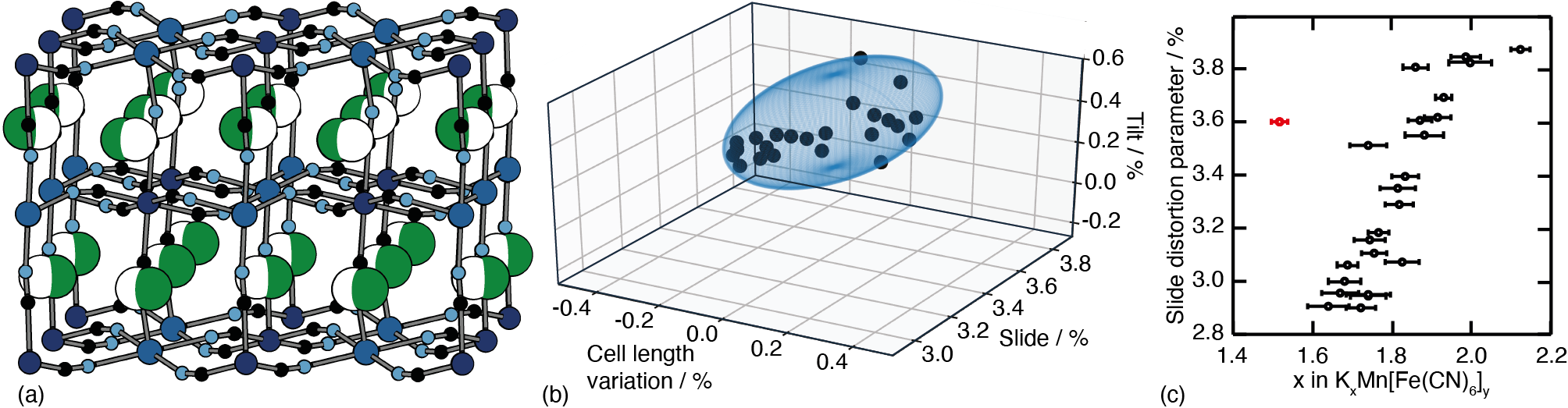}
	\caption{A-site slide distortions in monoclinic PBAs. (a) In K$_2$Mn[Fe], the K$^+$ ions (green spheres) are displaced from the aristotypic (high-symmetry) A-sites (white spheres). These displacements induce a local polarisation that couples such that successive layers `slide' in opposite directions. (b) The unit-cell parameters $a,b,c,\beta$ of the monoclinic PBAs K$_x$Mn[Fe]$_y$ can be used to calculate the three fundamental symmetry-adapted strains involved in the distortion away from cubic symmetry. The strain associated with A-site slides is an order of magnitude more significant than the two other strains, and varies throughout the compositional series. The ellipsoid illustrates the spread of data points and is a guide to the eye. (c) Other than for a single outlier (shown in red), the magnitude of A-site slide strain correlates well with K$^+$-ion concentration.\cite{Fiore2020} Further details and discussion are given in the SI.}
	\label{monoclinic}
\end{figure*}

One of the less obvious crystallographic consequences of tilts is that they allow the {\sf A}-site cations to off-centre. Doing so can enhance further the interaction with neighbouring cyanide ions, whilst also reducing cationic repulsion. Off-centering necessarily creates an effective dipole in each cavity, and one might expect that, at sufficiently high {\sf A}-site occupancies, the dipoles would interact in a collective way. This seems to be the case in the monoclinic structure of K$_2$Mn[Fe] (a relatively common structure type for PBAs\cite{Bie2017}): the dipoles are large\cite{Fiore2020} and arrange themselves in precisely the manner expected for interacting dipoles on the cubic lattice\cite{Allen2021} [Fig. \ref{monoclinic}a]. We refer to this collective displacement as a `slide' degree of freedom (to mirror the language of tilts\cite{Glazer1972} and shifts\cite{Bostrom2016} in perovskites and their analogues). 

Just how significant is this degree of freedom in other monoclinic PBAs? Here crystallography helps because the slide distortion has a specific signature in the monoclinic unit cell dimensions. In a technical sense, there are three contributions to the monoclinic distortion: one is this slide component, a second relates to a tilt distortion, and the third captures a variation in the unit cell lengths (see SI for further discussion.) Using crystallographic data from a recent study of 24 different K$_x$Mn[Fe]$_y$ samples\cite{Fiore2020} we find the slide distortion to be at least an order of magnitude more important than the other two (see Fig. \ref{monoclinic}b). Moreover the variation in slide distortion scales with potassium content (see Fig. \ref{monoclinic}c). The same analysis of other monoclinic PBAs yields similar results, and we provide a more comprehensive survey as Supporting Information (SI). The key result is that {\sf A}-site slides are most important in low vacancy, high alkali-metal ion content PBA systems.\cite{Xu2019}

On activation of an {\sf A}-site slide, the {\sf A}-site cation is trapped in a lower symmetry site with a likely steeper potential energy well. The pore-size itself is smaller too, which in turn is likely to affect ion mobility. Both factors will inhibit diffusion kinetics of the {\sf A}-site cation.\cite{Moritomo2009} Furthermore, if the aim is to maximise the {\sf A}-site cation content for a higher specific capacity active material then structural phase changes during electrochemical cycling will result in strains of the order of a few percent.

Just as tilts have a temperature dependence, so too does the degree of {\sf A}-site sliding decrease at higher temperatures. For example the monoclinic distortion vanishes altogether at 400K in Na$_{1.96}$Cd[Fe]$_{0.99}$ where the system returns to the parent cubic structure.\cite{Moritomo2021} It follows that such phase transitions occur at lower temperatures when the {\sf A}-site occupancy is reduced, since the magnitude of the dipole is smaller.\cite{Moritomo2011} In fact, phase transitions can be observed by changing the A-site occupancy alone\cite{Moritomo2009a} --- an extreme version of the compositional dependence discussed above.

The phase change is a result of a cooperative distortion, something we don't see in higher vacancy systems, because they can't reach a high enough {\sf A}-site concentration.\cite{Matsuda2012} A similar theme emerges when Jahn-Teller active metals occupy the {\sf P}-site.

\section{Jahn-Teller distortions}
The Jahn-Teller (JT) theorem states that any non-linear molecule with a spatially degenerate electronic ground state \emph{will} undergo a geometric distortion.\cite{Jahn1937} In the context of PBAs it's typically the {\sf P}-site metals which are JT-active; \emph{e.g.} Mn$^{3+}$ or Cu$^{2+}$.\cite{Moritomo2002,Chapman2006} Even in the presence of local JT distortions, however, the PBA structure may itself remain undistorted on average as a result of the presence of vacancies and/or the many other mechanisms that impart flexibility.\cite{Ojwang2016,Bostrom2019} 

\begin{figure}[b]
	\centering
	\includegraphics{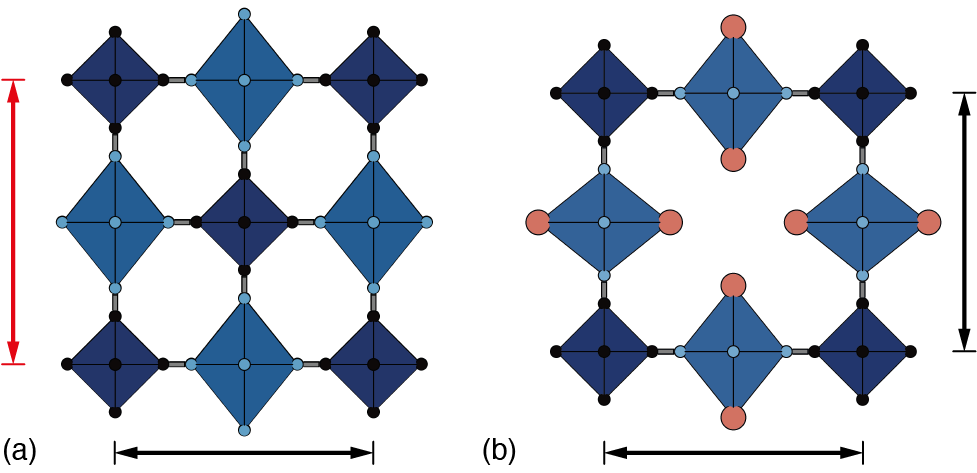}
	\caption{Interplay amongst cooperative JT order, hexacyanometallate vacancies, and lattice strain. (a) In low-vacancy PBAs such as Cu[Pt], the axial distortion of Cu$^{2+}$ coordination environments (light blue octahedra) is strongly coupled, resulting in a tetragonal strain. Here, for example, the cell axis denoted by the red arrows aligns with the JT axis and is longer than the orthogonal (black) axis. (b) The incorporation of hexacyanometallate vacancies allows for the JT distortion axis to vary throughout the crystal such that a cubic metric is retained on average. In this schematic, the central vacancy allows the JT axis to switch between vertical and horizontal orientations.}
	\label{JT}
\end{figure}

A good example of the effect of vacancies on collective JT order is given by the pair Cu[Pt] and Cu[Co]$_{\nicefrac{2}{3}}$. Stoichiometric Cu[Pt] has tetragonal symmetry owing to the cooperative JT distortion from Cu$^{2+}$ ions.\cite{Buser1974} Vacancies typically disrupt long-range JT order; hence the more familiar cubic structure seen in Cu[Co]$_{\nicefrac{2}{3}}$.\cite{Bostrom2019} A combination of single-crystal X-ray diffuse scattering measurements and Monte Carlo simulations have shown that the arrangement of vacancies can be predicted based upon crystal-field effects with Cu[Co]$_{\nicefrac{2}{3}}$ showing a stronger preference for centrosymmetric Cu geometries with vacancies aligned trans to one another.\cite{Simonov2020} This tendency means that JT-driven axial distortion can be accommodated by aligning the longer bonds with the vacancies to coordinate with structurally flexible water rather than the cyanide framework [Fig. \ref{JT}]. Once again, vacancies are disrupting connectivity, meaning distortions are local and not long-range; hence there is no splitting of Bragg reflections in the corresponding powder diffraction patterns.\cite{Ojwang2016} The problem of the long-range symmetry breaking in dilute JT systems has been extensively studied in conventional perovskites such as KCu$_{1-x}$M$_x$F$_3$ (M = Mg, Zn).\cite{Tanaka2005,Ghigna2010}

Returning to our leitmotif of the K$_x$Mn[Fe]$_y$ system, collective JT order has a strong effect on its performance as a cathode material. The key problem is that the low vacancy ($y\rightarrow1$) system in the fully oxidised state ($x\rightarrow0$) is tetragonal, as a consequence of the JT-active high-spin Mn(III) ($t_{2g}^{\,3}$\,$e_{g}^{\,1}$) configuration.\cite{Matsuda2011} During discharge, the Mn(III) is reduced to JT-inactive Mn(II) and cooperative JT order dissolves. There are a number of implications of JT distortions in this system. First, capacity loss is caused by structural damage to the cathode from repeated cycling between cubic and tetragonal phases.\cite{Jiang2019} Second, the strong vibronic coupling required for electron transfer at the Mn$^{3+}$ site limits rate capability when JT distortions are strongly coupled (this is the effect referred to as `JT-release' in Ref.~\citenum{Moritomo2013}). Third, the JT distortion lowers the reduction potential itself.\cite{Moritomo2013} Hence, addressing each of these issues will be key to optimising the electrochemical performance of this material, a point to which we will return in due course. 

\section{A-site cations}

The nature, position, and spatial ordering of {\sf A}-site cations in PBAs also has a large effect on their properties.

The impact of A-site cation type on the redox potential of the {\sf P}- and {\sf R}-site metals is well exemplified by the {\sf A}$_x$Co[Fe]$_y$ system.\cite{Bordage2020} Cation exchange of Na$^+$ for K$^+$ in Na$_{0.31}$Co$^{\rm{II}}$[Fe$^{\rm{III}}$]$_{0.77}$ drives a change in charge order to give K$_{0.31}$Co$^{\rm{III}}_{0.77}$Co$^{\rm{II}}_{0.23}$[Fe$^{\rm{II}}$]$_{0.77}$.\cite{Sato1997} Although the effect of changing {\sf A}-site ion is clear, the physical origin of this change in charge order is not well understood.\cite{Sato1997} Drawing on our analysis above, we expect the smaller Na$^+$ to interact more strongly with surrounding cyanide ions hence promoting framework distortion and varying the {\sf P}--N$\equiv$C--{\sf R} angles. Hence the crystal-field stabilisation energies at {\sf P}- and {\sf R}-sites should change as Na$^+$ is exchanged for K$^+$. Furthermore, it's understood that Na$^+$, unlike K$^+$, can sit in the square windows of the anionic framework,\cite{Ling2013} hence interacting more strongly with the cyanide ligand at the expense of the crystal field of Co. The combination of these effects would reduce the magnitude of the Co crystal-field splitting in the Na$^+$-containing system, so stabilising the Co$^{\rm{II}}$($t_{2g}^{\,5}$\,$e_{g}^{\,2}$)--NC--Fe$^{\rm{III}}$($t_{2g}^{\,5}$) ground-state configuration. 

Unlike the transition-metal octahedra, which are essentially fixed on the {\sf B} and {\sf B}$^\prime$ sites, the {\sf A}-site cations are in principle free to displace across a valley of extra-framework sites, with the particular site adopted depending both upon the ion/nanopore size, A-site type, and its concentration.\cite{Xiao2015} Traditional arguments based upon cation size can be used to explain which site is occupied, with only those bigger than Na$^+$ consistently occupying the largest interstitial site at the centre of the nanopore.\cite{Ling2013} Occupancy of this site comes at the expense of electrostatic interaction with the cyanide ligands; hence we observe tilts and slides --- as discussed in the previous sections --- in PBAs with high K-ion concentrations.\cite{Bie2017}

\begin{figure}
	\centering
	\includegraphics[width=\columnwidth]{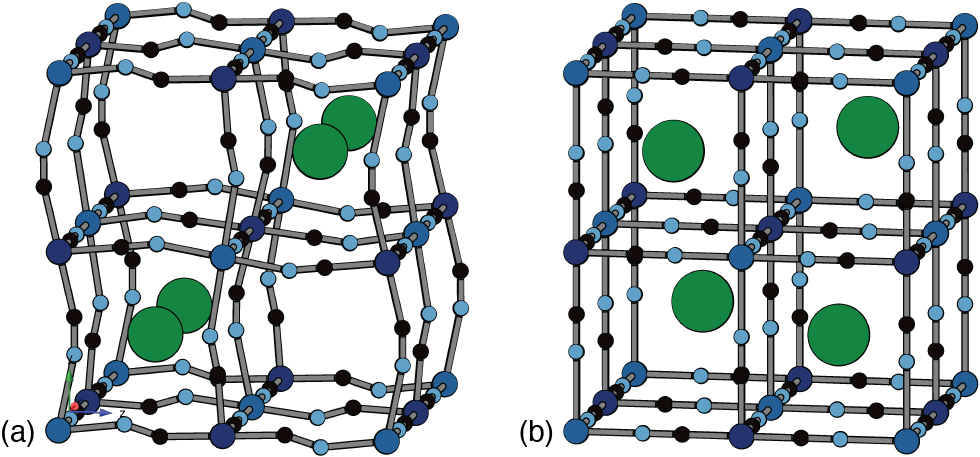}
	\caption{Two representative types of long-range cation order in PBAs. (a) In Rb$_{0.85}$Cu[Fe]$_{0.95}$, correlated tilts give rise to two symmetry-distinct extraframework channels; the Rb$^+$ ions (shown as green spheres) preferentially occupy the narrower channels so as to maximise electrostatic interactions with the anionic framework. (b) In Cs$_{0.97}$Cu[Fe]$_{0.99}$, the absence of any long-range tilt distortions means that all channels are equivalent by symmetry. The Cs$^+$ ions (green spheres) now occupy alternating {\sf A}-sites so as to minimise electrostatic repulsion between one another. Figure adapted from Ref.~\citenum{Matsuda2012a}.}
	\label{A-site}
\end{figure}

Positional ordering of {\sf A}-site cations has been observed in Rb- and Cs-containing PBAs [Fig.~\ref{A-site}]. Cation ordering can reduce repulsions and increase electrostatic attraction with the cyanide ligands.\cite{Liu2019} It can be affected also by the reduction in symmetry associated with other degrees of freedom. For example, in Cs$_{0.97}$Cu[Fe]$_{0.99}$ the Cs-ions fit well in the {\sf A}-site cavity and adopt a `dot-like' arrangement\cite{Matsuda2012a} ($R$-type order) to minimise repulsions. In Rb$_{0.85}$Cu[Fe]$_{0.95}$, the activation of tilts divides the {\sf A}-site cavities into two types, one of which is preferentially occupied by the smaller Rb ions forming a `rod'-like arrangement ($M$-type order).\cite{Matsuda2012a} Although harder to characterise using X-ray diffraction for the lighter alkali metal ions, their ordering and corresponding impact on lattice strain would be helpful to understand when it comes to working with PBA electrodes, as is the case for Li-ion cathodes.\cite{Wang1999} Preliminary computational studies have shown cation ordering should be prevalent in PBAs.\cite{Liu2019}

\section{Hexacyanometallate vacancies}

It is difficult to decouple the dual effects of [{\sf R$^n$}(CN)$_6$]$^{(6-n)-}$ vacancies and {\sf A}-site cations, since as the number of vacancies is reduced, {\sf A}-site cations are necessarily intercalated to charge-balance. In the {\sf A}$_x$Co[Fe]$_y$ system discussed above, for example, it is not only the A-site cations but also the vacancies that affect crystal-field stabilisation energies --- because increasing hexacyanoferrate content means Co--OH$_2$ bonds are replaced by Co--NC bonds. Hence we observe Co$^{\rm{II}}$[Fe$^{\rm{III}}$]$_{0.67}$ and Co$^{\rm{III}}$[Fe$^{\rm{II}}$]$_{0.97}$ to be the dominant configurations across the Cs$_x$Co[Fe]$_y$ family.\cite{Bleuzen2000} There is a functional consequence of these different electronic configurations. Whereas the former is a candidate for photoinduced charge transfer, the latter is not: it lacks sufficient flexibility to accommodate the associated change in bond lengths.\cite{Escax2003} 

These dual effects of vacancies --- namely, variation in both crystal-field stabilisation energies and framework flexibility --- seem impossible to ignore when seeking to reduce the number of vacancies in K$_x$Mn[Fe]$_y$.\cite{Bie2017} Reducing vacancy concentration should improve capacity, but will do so at the expense of the framework flexibility that improves cycle life.\cite{Wessells2011} Hence, one anticipates some compromise must be found\cite{Hosaka2021} --- in which case we have to consider also the potential for vacancies themselves to order.

\begin{figure*}
	\centering
	\includegraphics[width=\textwidth]{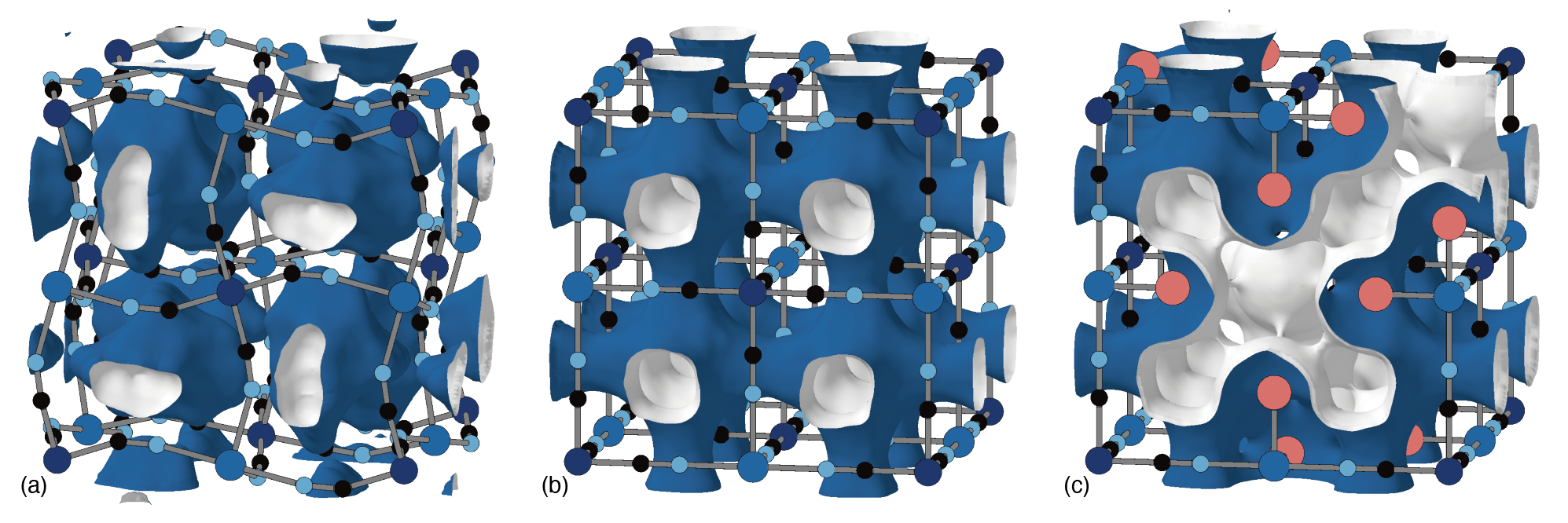}
	\caption{Effect of structural complexity on pore-networks in PBAs. (a) The activation of correlated tilt distortions, such as for the monoclinic PBA shown here, isolates the cavities centred on each {\sf A}-site. Here the void volume is shown as isosurfaces with a blue exterior and white interior; note that neighbouring volumes do not connect. (b) In the aristotypic structure type, these cavities connect \emph{via} narrow windows at the cube faces to form a pore-network structure with simple cubic topology. (c) The incorporation of hexcyanometallate vacancies dramatically opens up the pore-network, which now includes channels that run along the cube face diagonals. These channels are necessarily lined by the water molecules (red spheres) that cap the vacant coordination sites of neighbouring transition-metal cations. All isosurfaces are generated using the same probe radius.}
	\label{vacancies}
\end{figure*}

The tendency for vacancies to order was established in Prussian blue itself (Fe$^{\rm{III}}$[Fe$^{\rm{II}}$]$_{0.75}$), where they were found to avoid one another.\cite{Buser1977} It's now known that vacancies are strongly correlated --- if not long-range ordered --- across a wide range of {\sf P}$^{\rm{II}}$[{\sf R}$^{\rm{III}}$]$_{\nicefrac{2}{3}}$ PBAs.\cite{Simonov2020} The form of these vacancy correlations primarily depends on the nature of the {\sf P}-site metal and the particular synthesis conditions used. Since it has been demonstrated that larger cations --- Rb$^+$ and possibly K$^+$ --- preferentially migrate \emph{via} vacancy channels, the vacancy content and pore-network geometry has important consequences for diffusivity in a K$_x$Mn[Fe]$_y$ electrode.\cite{Moritomo2009} The nature of pore-networks in PBAs and the interplay with structural distortions and hexacyanometallate vacancy distributions are illustrated in Fig.~\ref{vacancies}.

\section{Hydration}
It would be easy to view water as a spectator in PBA chemistry since the compounds are so insoluble, but it is now clear that the hydration environment of a PBA influences both structural complexity and functionality.\cite{Kareis2012,Ojwang2021} It has long been known that there are essentially two distinct kinds of water in PBAs: ligand water, which bonds tightly to the {\sf P}-site metals that neighbour a hexacyanometallate vacancy, and interstitial water which occupies the nanopore.\cite{Herren1980} When ligand water neighbours an interstitial water molecule it moves `off-axis' to form a hydrogen bond, but if the interstitial water is removed by heating, then the off-axis water appears to return to the standard ligand water site.\cite{Kim2009} 

In the case of Li$^+$ and Na$^+$, which don't necessarily occupy the body centre of the nanopore, interstitial water can accompany the {\sf A}-site cation\cite{Zhou2020} to which it is strongly bound.\cite{Liu2021} The hydrophilicity of the Li- and Na-ions raises the Gibbs free energy of insertion, which is why one finds lower Na$^+$ occupancy compared with K$^+$.\cite{Zhou2020} Conversely, larger cations replace interstitial water: hence the limiting stoichiometry K$_2$Mn[Fe] is anhydrous.\cite{Deng2021}

Since interstitial water is relatively labile on heating, the drying conditions used in PBA synthesis can by itself alter their structure.\cite{Song2015} For example, the system Na$_2$Mn[Mn] switches between an open monoclinic form and a dense rhombohedral form with the loss of two formula units of H$_2$O [Fig. \ref{tilts}].\cite{Kareis2012} Sodium-ion PBAs are in general particularly sensitive to drying conditions, with high temperature vacuum drying favouring a dense rhombohedral structure,\cite{Brant2019} in the same way that PBAs collapse under external pressure.\cite{Bostrom2019} These structural transitions effect the crystal-field stabilisation energy of the {\sf P}- and {\sf R}-site metals. In the dehydrated Na$_{1.89}$Mn[Fe]$_{0.97}$ system the competing ionisation energy and crystal-field stabilisation energy mean the redox potentials of Fe$^{3+/2+}$ and Mn$^{3+/2+}$ overlap to produce a single plateau in the charge/discharge profile, distinct from the two plateaux observed for the hydrated system.\cite{Wu2017} 

\begin{figure}[b]
	\centering
	\includegraphics{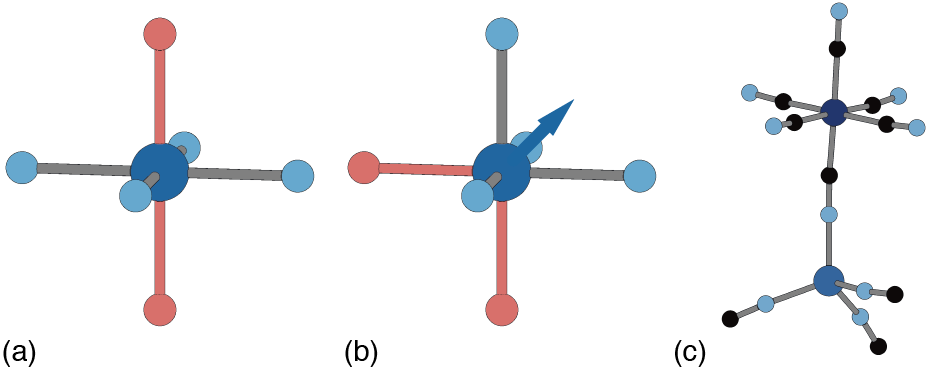}
	\caption{{\sf P}-site hydration geometries and their implications. (a) The \emph{trans}-[{\sf P}N$_4$O$_2$] configuration, \emph{e.g.}\ as favoured by Cu$^{2+}$, is structurally stable to dehydration since both hydrated and dehydrated coordination geometries are non-polar and so constrain the {\sf P}-site position. (b) By contrast, the \emph{cis}-[{\sf P}N$_4$O$_2$] configuration observed in some Zn- and Mn-containing PBAs has C$_{2\rm v}$ symmetry. On dehydration, the {\sf P}-site cation displaces along the local C$_2$ axis (shown here as an arrow) such that its coordination geometry becomes approximately tetrahedral. (c) A fragment of the network structure of rhombohedral (non-PBA) polymorph of Zn$_3$[Co(CN)$_6$]$_2$, in which Zn$^{2+}$ ions are tetrahedrally coordinated by four hexcyanocobaltate anions.\cite{Rodriguez2007} The geometry at the Zn$^{2+}$ site is similar to that achieved on dehydration of the \emph{cis}-N$_4$O$_2$ coordination environment shown in (b).}
	\label{hydration}
\end{figure}

Ligand water is more strongly bound to the PBA lattice, but its removal can also have significant structural and electronic consequences. If there are two vacancies surrounding the {\sf P}-site transition-metal such that its octahedral coordination is [{\sf P}N$_4$O$_2$] then the waters can arrange either \emph{cis} or \emph{trans}. In a \emph{trans} arrangement, removal of bound water leaves the {\sf P}-site in a (geometrically stable) D$_{4\rm{h}}$ environment, but removal of \emph{cis} water-pairs leaves the {\sf P}-site in a polar C$_{2\rm{v}}$ arrangement. The {\sf P}-site metal necessarily relaxes along the polar C$_2$ axis to acquire tetrahedral coordination, which in turn affects both structure and electronics [Fig \ref{hydration}].\cite{Sato2003}

Ligand water becomes reactive at high cell potentials, causing unwanted side reactions with the electrolyte,\cite{Brant2019} exposing the {\sf P}-site metal to dissolution, and leading to capacity loss.\cite{Fiore2020} Given that ligand water cannot be removed entirely by drying before cell assembly, the only viable method for excluding it from the PBA structure seems to be to replace it with hexacyanometallate anions.\cite{Zhou2019} This strategy must be tensioned against the potential benefits of some vacancy inclusion, as discussed above.

\section{K$_{\textbf{x}}$Mn[Fe]$_{\textbf{y}}$: a truly complex cathode material}

To illustrate the functional implications of the various types of structural complexity covered in our Focus article, we conclude by returning to our core example, K$_x$Mn[Fe]$_y$. We have already seen that this system displays many of the complexities covered in our article, and we now argue that understanding the interplay of these various aspects will be key to optimising its performance as a K-ion cathode material. We consider in turn the two key strategies currently proposed in the literature: namely, eradicating vacancies altogether,\cite{Deng2021} and actively including vacancies in concentrations as large as 15\%.\cite{Hosaka2021}

The first strategy is based on the assumption that, by removing vacancies from the structure, one can at once both maximise capacity and remove water entirely from the framework.\cite{Deng2021} Doing so would solve the problem of side reactions in a non-aqueous electrolyte at high potentials\cite{Zhou2019} driving dissolution of the active material and causing capacity fade.\cite{Fiore2020} The drawback of this strategy is that there remain two significant phase changes on cycling,\cite{Jiang2019} and the reduced vacancy concentration lowers ionic conductivity.\cite{Hosaka2021} Taken together, these two points mean the theoretical capacity is very hard to achieve in practice, particularly at high rates.\cite{Dhir2020} 

Doping on the {\sf P}-site is an obvious strategy to overcome the tetragonal distortion caused by JT-active Mn$^{3+}$ in low vacancy Mn$^{\rm{III}}$[Fe$^{\rm{III}}$]$_y$. As little as 12\% Ni doping in sodium-ion Mn[Fe]$_{0.98}$ was enough to prevent tetragonal distortion.\cite{Yang2014} Avoiding global distortion to the structure activates the Mn reduction reaction (Mn$^{3+}$\,/\,Mn$^{2+}$), raising the specific energy.\cite{Moritomo2016} The drawback is that substitution with Ni does sacrifice some capacity since it is electrochemically inert, but it seems to give some security to the framework, improving cycling stability.\cite{Yang2014} Such a loss in capacity can be avoided by doping with electrochemically-active Fe$^{2+}$. One third doping in K$_x$Fe$_{0.33}$Mn$_{0.67}$[Fe]$_{0.98}$ suppresses the tetragonal distortion, also improving cycling stability.\cite{Jiang2019} Furthermore, the rate capability is improved by decreasing the band gap and lowering the K-ion diffusion activation energy.\cite{Jiang2019} So it seems {\sf P}-site doping is an excellent strategy to supress the JT distortion, but what about the monoclinic distortion? It doesn't appear to be avoidable in the same way, but the evidence suggests the highly crystalline samples with a strongly cooperative monoclinic distortion show improved electrochemical performance.\cite{Zhou2019,Xu2019} If this is the case, then the distortion should be studied further to gain a fundamental understanding that can better inform the design of low vacancy K$_x$Mn[Fe]$_y$.

If the optimal solution is a PBA that includes vacancies for superior rate capability and structural stability,\cite{Hosaka2021,Hurlbutt2018} then the biggest problem is the presence of water in the structure.\cite{Zhou2019} As well as careful choice of electrolyte to minimise damaging reaction with water,\cite{Dhir2020} alterations should be made to the structure to prevent water from escaping into the cell, since it appears impossible to remove prior to assembly.\cite{Song2015} Coating the particles could certainly be a promising solution since it also prevents dissolution of Mn ions.\cite{Feng2021,Gebert2021} Vacancy network correlation could also play a role since they appear to improve ionic conductivity.\cite{Moritomo2009,Hosaka2021} 

\section{Concluding remarks}

If there is one take-home message from our brief survey it is surely that PBAs support an extraordinarily rich diversity of structural complexity. Based on the collective experience of studying similarly complex families --- \emph{e.g.}\ the manganite perovskites, where the interplay of charge, spin, lattice, and orbital degrees of freedom gives rise to a variety of unexpected and important physical properties (\emph{e.g.}\ colossal magnetoresistance) --- one expects that learning both how to control the various individual distortion types we discuss here and how to exploit the interactions between different distortions will together be crucial for developing the next generation of functional PBA materials. Hence there is a clear motivation for fundamental structural studies that systematically explore the vast parameter space accessible experimentally to PBAs.\cite{Fiore2020}

While our emphasis here in terms of applications has been on the use of PBAs as battery materials, there is every reason to expect that structural complexity plays an important role in the many other fields in which PBAs find application. We have already highlighted one or two examples in the context of photosensitivity and charge order, but the implications for spin-crossover,\cite{Kosaka2005a,Bostrom2020b} proton conductivity,\cite{Ohkoshi2010} gas storage,\cite{Kaye2005} and cooperative magnetism\cite{Ohkoshi2005,Ferlay1995} are all straightforward to envisage.

Looking forward, one of the most difficult challenges to be overcome is for the community to develop a robust understanding of the \emph{local} structure of high-symmetry PBAs, and the mechanistic importance of local distortions for function. In the various examples given here, we have (understandably) relied on long-range symmetry breaking to identify the structural degrees of freedom at play. Yet the intuition is that these same degrees of freedom are just as important in the absence of long-range symmetry breaking. In related fields, such as that of the disordered rocksalt cathode materials, computation is crucial in developing an atomistic picture of local structure.\cite{Ji2019,Clement2020} However, the more open framework structure of PBAs relative to the dense lattice of rocksalt oxides, the importance of H$_2$O, and that of vacancies and their correlations collectively mean that the computational challenge here is substantial indeed. The development of efficient force-fields, a current priority in the field of metal--organic frameworks,\cite{Vanduyfhuys2015,Durholt2019,Rogge2021} may be particularly useful in allowing simulation on the requisite $\sim10$\,nm scale. Experimentally, there is an obvious motivation for further systematic local-probe investigations, exploiting \emph{e.g.}\ total scattering, nuclear magnetic resonance spectroscopy, Mössbauer spectroscopy, and/or X-ray absorption spectroscopy.

Will it be worth it? We argue, emphatically, yes: despite the obvious challenges posed by the many sources of complexity in PBAs, this diversity of structural, compositional, and electronic degrees of freedom presents an almost unparalleled opportunity to exploit complexity in functional materials design.

\section*{Conflicts of interest}
There are no conflicts to declare.

\section*{Acknowledgements}
We gratefully acknowledge financial support from the E.R.C. (Grant 788144), the ISCF Faraday Challenge projects SOLBAT (grant number FIRG007) and LiSTAR (grant number FIRG014) as well as the Henry Royce Institute (through UK Engineering and Physical Science Research Council grant EP/R010145/1) for capital equipment. The authors would like to thank Hanna Boström (Stuttgart), Arkadiy Simonov (Zurich), Chris Howard (Newcastle), Trees De Baerdemaeker (BASF), and Samuel Wheeler (Oxford) for many useful discussions.

\bibliography{mh_2021_pbas} 

\end{document}